\documentclass[pageno]{jpaper}

\newcommand{\ICTTechnicalReportnumber}{20180310}

\usepackage[normalem]{ulem}

\usepackage{authblk}

\usepackage{amsmath}

\usepackage{amsthm}

\usepackage{url}

\usepackage{graphicx}

\newenvironment{prf}{\noindent {\it Proof}  \ }{\hfill}

\newcommand{\tabincell}[2]{\begin{tabular}{@{}#1@{}}#2\end{tabular}}

\begin{document}

%\title{Gene-Pattern: Breaking the Bound between General-purpose and Special-purpose\\
%\Huge Should we customize architecture for each application?}
\title{Gene-Patterns: Should Architecture be Customized for Each Application?}

\author[*]{Yuhang~Liu}

\author[*]{Luming~Wang}

\author[*]{Xiang~Li}

\author[+]{Yang~Wang}

\author[*]{Mingyu~Chen}

\author[*]{Yungang~Bao}

\affil[*]{Institute of Computing Technology, Chinese Academy of Sciences}

\affil[+]{Shenzhen Institutes of Advanced Technology, Chinese Academy of Sciences}

\date{}
\maketitle
\let\thefootnote\relax\footnotetext{This paper published in arxiv was the exact version we submitted to asplos'18 two years ago.}
\thispagestyle{empty}

\begin{abstract}

Providing architectural support is crucial for newly arising applications to achieve high performance and high system efficiency. Currently there is a trend in designing accelerators for special applications, while arguably a debate is sparked whether we should customize architecture for each application. In this study, we introduce what we refer to as \emph{$Gene\text{-}Patterns$}, which are the base patterns of diverse applications. We present a Recursive Reduce methodology to identify the hotspots, and a HOtspot Trace Suite (HOTS) is provided for the research community. We first extract the hotspot patterns, and then, remove the redundancy to obtain the base patterns. We find that although the number of applications is huge and ever-increasing, the amount of base patterns is relatively small, due to the similarity among the patterns of diverse applications. The similarity stems not only from the algorithms but also from the data structures. We build the Periodic Table of Memory Access Patterns (PT-MAP), where the indifference curves are analogous to the energy levels in physics, and memory performance optimization is essentially an energy level transition. We find that inefficiency results from the mismatch between some of the base patterns and the micro-architecture of modern processors. We have identified the key micro-architecture demands of the base patterns. The Gene-Pattern concept, methodology, and toolkit will facilitate the design of both hardware and software for the matching between architectures and applications.

\end{abstract}

\section{Introduction}

In its seventy-year history, computer science at present is at a turning point, which is largely manifested in three aspects. First, Moore's law is deemed to be approaching to its end, as the increasing of the density of chip components is gradually slowing down and significantly deviating from the forecast of Moore's law~\cite{Mitch2016Moore}. It seems that the era of post Moore's law is coming. Second, brand new applications are emerging daily, which not only aggravates the memory wall problem but also requires new underlying micro-architectures, especially for the memory systems~\cite{Kouzes2009The}. Third, transistor scaling and voltage scaling are no longer in line with each other, which results in the utilization wall and the failure of Dennard scaling~\cite{Hardavellas2011Toward}, and thus calls for an architecture providing high hardware utilization and high energy efficiency.

Applications specify the demands for processor architecture. However, there exists a mismatch between the existing processor architecture and the increasingly diverse applications. General-purpose CPUs and GPUs are facing difficulties adapting to the application diversity. As a result, for a specified application, a large fraction of transistors could be wasted, and thus only a small fraction of peak performance is achieved. Although FPGAs provide elasticity when they are used to customize the architecture for applications, the customization process is very time consuming. As a result, FPGAs have not been used as widely as CPUs and GPUs.

To take advantage of the three components (i.e., CPUs, GPUs and FPGAs), architectures have evolved towards heterogeneous chip multi-processors (CMPs) that comprise a mix of cores and accelerators, and thus provide some architecture diversity for applications. In recent years, the heterogeneous CMPs have been customized for special domains such as those in machine learning related applications~\cite{conf/asplos/LiuCLZZTFZC15}. These applications are pervasive but they only represent a small portion of all the applications available today and in future. Compared to the increasing application diversity, the diversity in heterogeneous CMPs lags far behind. Therefore, a prominent issue is whether we should customize the CMP architectures for each application. In this study, our answer is that: \emph{not necessary}!

Our solution in this study is inspired by rethinking the knowledge in biology. In biology, it is well known that the properties of lives are determined by the combination of a series of genes. If we think of applications as ``lives'', then all we need to do is to find the corresponding ``genes''. Once we find the ``genes'' of applications, we only need to focus on the genes, the types of which are very limited. In that sense, the design process of computing systems can benefit from ``genetic engineering''. 

%In chemistry, the periodic table of elements reveals the building blocks of matter. If we think of applications as ``matter'', then all we need to do is to find the corresponding ``elements''. Once we find the ``elements'' of applications, we need only focus on the elements, the types of which are very limited. In that sense, the computing system design process can benefit from ``chemical engineering''.

This paper provides the first (to the best of our knowledge) comprehensive study on the hotspot characteristics of quite diverse applications. Specifically, we have examined the patterns of 106 hotspots of 62 benchmarks from eight representative suites, including SPECint and SPECfp from SPEC CPU2006~\cite{spradling2007spec}~\cite{Henning2006SPEC}, PARSEC~\cite{Bienia2008The}, BigDataBench~\cite{Wang2014BigDataBench}, MLPack~\cite{Curtin2012MLPACK}, HPCC~\cite{Luszczek2006The}, HPCG~\cite{Heroux2013HPCG} and Graph500~\cite{Graph500}. The applications are from different domains from big data analytics to high performance computing, from single-thread to multi-thread. Our study reveals lots of interesting findings and provides useful guidance for memory access pattern detection, classification, and optimization. 

In this study, for the first time, memory access patterns have been experimentally identified, mathematically formalized, and graphically visualized simultaneously for diverse applications. This pattern study reveals the essence of memory performance bottleneck and it is useful for code optimization and architecture design. The base patterns give incentives to computer system designers to invest in capabilities that will impact the collective performance of these essential patterns. 

In this study, our main contributions are the following:

\begin{itemize}
\item We propose a Recursive Reduce (ReRe) methodology for identifying and representing the memory access pattern of diverse applications. Accompanying this, we also provide a HOtspot Trace Suite (HOTS) for the research community.
\item We compare the similarities and differences among the programs in different domains. We propose a method using five metrics, Reuse aware Locality (RaL)~\cite{Liu2017CaL}, pipeline stall degree, L2 and beyond active degree, prefetch/request ratio and latency non-hidden degree. We consider metrics, codes and pattern figures simultaneously to explore insightful results.
\item We build the Periodic Table of Memory Access Patterns (PT-MAP), where memory performance of a pattern is determined by the pattern's location in the table. The indifference curves in PT-MAP are analogous to the energy levels in physics, and memory performance optimization is essentially an energy level transition. 
\item We extract the ``genes'' of applications, which are called base patterns, gene-patterns or meta-patterns, defined as the minimal complete set (MCS) of building blocks of memory access patterns. We find that today's predominant micro-architecture cannot match all the base patterns, which thus calls for a series of targeted changes to micro-architecture in the fetching granularity, caching and prefetching policy. 
\end{itemize}

The remainder of this paper is organized as follows. Section~\ref{sec:rrm} describes the recursive reduce methodology. Section~\ref{sec:ch} characterizes the hotspots with metrics, and Section~\ref{sec:vp} visualizes the patterns and presents the base patterns. We discuss Gene-Pattern aware accelerators and programming in Section~\ref{sec:bg} and Section~\ref{sec:rg}, respectively. We summarize related work in Section~\ref{sec:rw} and conclude in Section~\ref{sec:Conclusions}.

\section{Recursive Reduce Methodology}
\label{sec:rrm}

\subsection{Overview}

The significance of understanding the characteristics of applications has been well recognized by research community. To design a desirable architecture, we first need a good understanding of application behaviors. Specifically, it is necessary for architects to explore the application space, which is the collection of all possible applications. However, the number of applications that currently exist and are likely to exist in future is countless. Furthermore, each application consists of many lines of code. What makes things worse is the extremely low speed of cycle-accurate simulation, which is 5 orders of magnitude slower than physical execution~\cite{binkert2011gem5}. As a result, exhaustive consideration of application-to-architecture mappings is infeasible.

\begin{figure}
	\centering
	\includegraphics[width=0.8\linewidth]{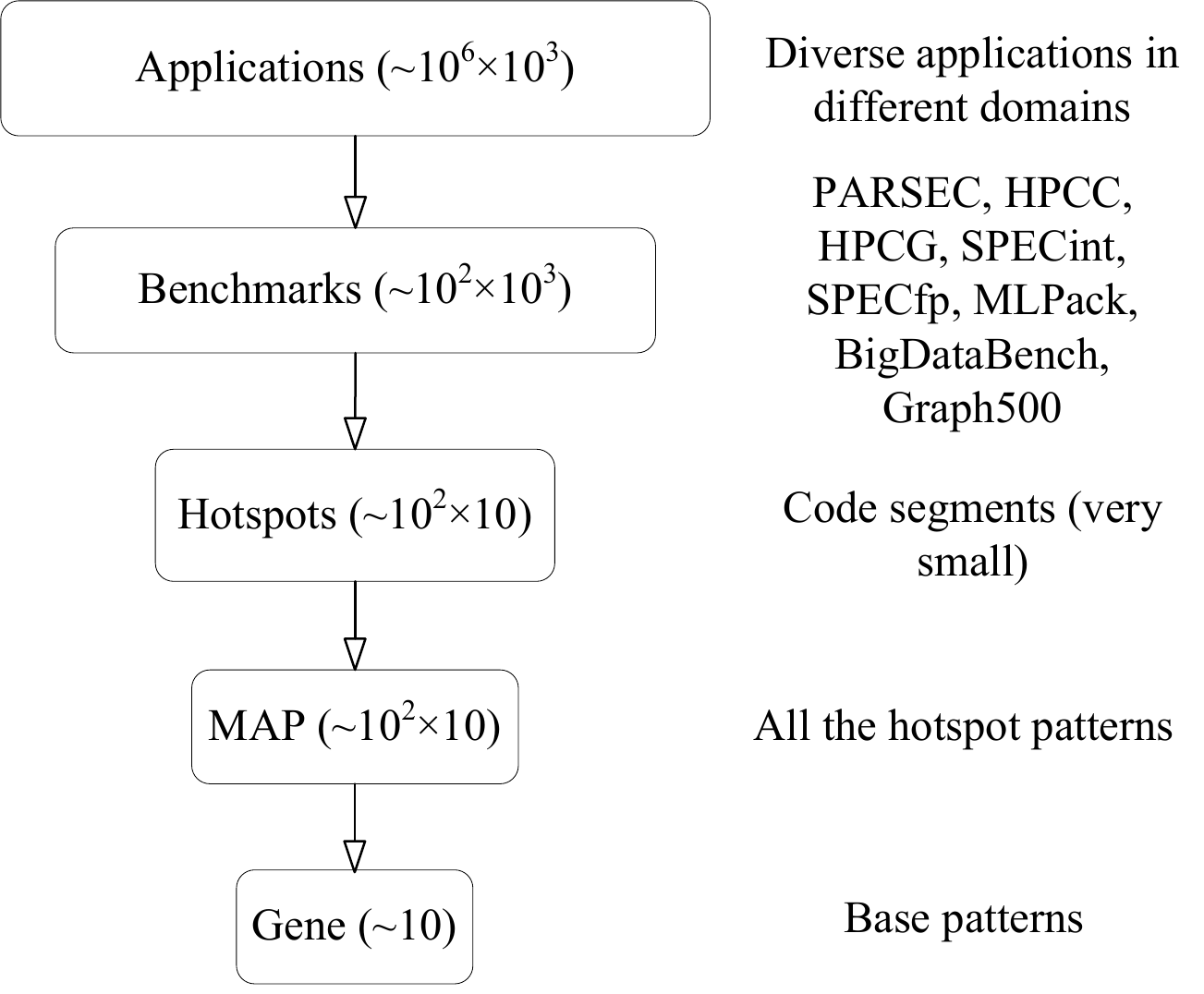}
	\caption{The Recursive Reduce methodology for building the base patterns of applications ($a\times b$ means that there exist $a$ different elements and each element is of size $b$)}
	\label{fig:1}
\end{figure}

In this study, we present a Recursive Reduce method (abbreviated as ReRe) for identifying and representing the patterns of diverse applications. As shown in Fig.~\ref{fig:1}, we use various benchmark suites in the first step as the representative of all the real applications. Previous studies have characterized applications from the perspective of the whole application; this seems a natural choice~\cite{Bienia2008The}~\cite{Wang2014BigDataBench}. However, the benchmark programs are still very large and difficult to analyze. We observed the fact that, in an application program, although there are many lines of code, usually only a very small portion consumes most of the execution time. Moreover, an application may comprise several hotspots, and one hotspot usually corresponds to a pattern, thus an application may have more than one pattern during its execution. If we measured the average value of metrics of the whole application, the diversity of the patterns in an application is hidden by the average. Therefore, in this study we characterize applications from the perspective of the identified hotspots rather than the whole application. 

Modern commodity processors provide a number of hardware performance counters to support micro-architecture level profiling. We use PAPI~\cite{Dongarra2001Using} to collect about 30 events, whose numbers and unit masks can be found in the Intel Developer's Manual~\cite{inteloptimize}. Based on the events, we use HPCToolkits~\cite{Adhianto2010HPCTOOLKIT} and Perfexpert~\cite{Burtscher2010PerfExpert} to identify the hotspots. 

To recognize hotspots, debug information produced by the compiler is required. The debug information includes the mapping between source code and the executable binary. An application's executable binary (embedded with debug information) is executed and profiled by HPCToolkits. Then, a performance database is produced, which contains an application's source code information and performance metrics that are calculated with the performance counter values. Finally, the tool Perfexpert is used to analyze the data in performance database in order to identify the code segments whose importance is great than a threshold (e.g., 5\%). A hotspot's \emph{importance} is computed in Eq.~\eqref{Importance-defination} as follows, 

\begin{align}
\label{Importance-defination}
Importance = \frac{t \times n}{t_s + n \times t_p} \textgreater \Delta\%,
\end{align}
where $t$ is the time taken by the code segment, $n$ is the number of threads, and $T_s$ and $T_p$ are the time taken by the sequential and parallel sections of the total benchmark, respectively. In this manner, the hotspots of an applicaion can be identified.

For each hotspot, we conduct detailed analysis using metrics, which are the attributes of the hotspot's pattern. This fine-grained analysis enables us to gain more insights, compared to a coarse-grained one. We then plot the access patterns and analyze the corresponding codes. Finally and most interestingly, with ReRe method, we hierarchically mine the essential patterns, based on which we remove redundancy and obtain the base patterns.

%To find the source code of a hotspot, the compiler must generate the debug information that is inserted into the final executable binary file. The debug information includes variables, source code, etc, which enables source-level debugging. By parsing the debug information that includes the performance counter values, the source code of each hotspot can be found. Thus, the debug options should be enabled when the programs are compiled. %However, currently there is a limitation for profiling the programs running in virtual machines. The programs that are executed in virtual machines, for instance, Java programs that are running in Java Virtual Machines (JVM), cannot be profiled. The reason is that the executable file is oriented to the virtual machine rather than the physical machine, the debug information cannot collect the performance counters of physical machine, and thus we cannot identify the hotspots and their source codes.

\subsection{Benchmark and Platform}
We use benchmarks from eight suites in which each benchmark suite is designed to exercise computational and memory access patterns that much closely match a different and broad set of important applications. Specifically,

\begin{itemize}
\item \emph{BigDataBench}~\cite{Wang2014BigDataBench} is for large footprint workloads, modelling typical big data application domains: search engine, social networks, e-commerce, multimedia analytics, and bioinformatics; 

\item \emph{MLPack}~\cite{Curtin2012MLPACK} is a collection of artificial and real-world machine learning workloads; 

\item \emph{PARSEC}~\cite{Bienia2008The} is a benchmark suite composed of multithreaded programs. The suite focuses on emerging workloads and was designed to be representative of next-generation shared-memory programs for chip-multiprocessors; 

\item \emph{SPECint}~\cite{spradling2007spec}~\cite{Henning2006SPEC} comprises single thread integer benchmarks from SPEC CPU2006, stressing a system's processor, memory subsystem and compiler; 

\item \emph{SPECfp}~\cite{spradling2007spec}~\cite{Henning2006SPEC} is similar to SPECint and is also a subset of SPEC CPU2006, but focuses on float point processing;

\item \emph{HPCC}~\cite{Luszczek2006The} includes $LINPACK$ and $RandomAccess$ and tests a number of independent attributes of the performance of high-performance computer (HPC) systems;

\item \emph{HPCG}~\cite{Heroux2013HPCG} is intended as a complement to the $LINPACK$ benchmark, currently used to rank the Top500 computing systems;

\item \emph{Graph500}~\cite{Graph500} is to augment the LINPACK with data-intensive applications. Because graph algorithms are a core part of most analytics workloads, Graph500 offers a forum for the community and provides a rallying point for data-intensive supercomputing problems.

\end{itemize}

We conduct our study on a typical commodity server with two Intel Xeon E5-2630 v4 and 64GB of DRAM in each blade. Each Intel E5-2630 processor includes ten aggressive out-of-order processor cores with a deep (three-level) on-chip cache hierarchy. Table~\ref{Tab:1} summarizes the architectural parameters of the experimental system.

%\begin{scriptsize}
\begin{table}[h!]
  %\scriptsize
\centering
\caption{The experimental system parameters}
\label{Tab:1}
\begin{tabular}{|c||m{0.35\columnwidth}| }
\hline  % \\ 
Processor & 14nm Intel Xeon E5-2630 v4 operating at 2.2 GHz\\ [0.5ex] 
\hline \hline
Chip number & 2 LGA sockets \\
\hline
Cores per chip & 10 OoO cores\\
\hline
Threads per core & 2\\
\hline
L1 dcache and icache latency & 4 cycles \\ 
\hline
L2 latency & 12 cycles \\ 
\hline
L3 latency & 40 cycles \\ 
\hline
Memory latency & 150$\pm$50 cycles \\ 
\hline
FP latency & 2 cycles \\
\hline
FP slow latency & 18 cycles \\
\hline
TLB latency & 45 cycles \\
\hline
Memory & 4 DDR3 channels, 16GB per channel\\
\hline
\end{tabular}
\label{table:Table} 
\end{table}
%\end{scriptsize}

We completely run the 62 benchmarks from eight different suites. Using the performance counters with the criterion in Eq.~\eqref{Importance-defination}, we have identified 106 hotspots, each having a small number of lines of code. On average, the importance of all the 106 hotspots is 23\%. In other words, although each hotspot only includes a small number of lines of code, it takes more than 20 percent of the total execution time. We also use PIN~\cite{Bach2010Analyzing} to collect the traces of each hotspot. The hotspot locations in the programs and the traces of the hotspots are formed into a HOtspot Trace Suite called \emph{HOTS}, which will be released for the research community.% after the publication of this work.

\section{Characterizing the Hotspots with Metrics}
\label{sec:ch}
In this section, we use metrics to characterize the hotspots. Specifically, for each hotspot, we evaluate the effectiveness of the following micro-architecture components: pipelines, on-chip cache hierarchy, prefetchers and memory controllers. Modern CMP comprises lots of processing cores in a single chip, and each core can issue several instructions in a cycle and execute instructions out of order (OoO). We propose \emph{pipeline stall degree} to quantify the pipeline efficiency. CMP has a deep (three-level) cache hierarchy, where each core is equipped with a split L1 instruction and data cache, and a unified L2 cache, and LLC is shared among all the on-chip cores. We propose an evaluation method using \emph{Reuse aware Locality (RaL)}~\cite{Liu2017CaL}, \emph{L2 and beyond active degree} and \emph{latency non hidden degree} to quantify the cache hierarchy effectiveness. To mitigate the latency gap between L2 and LLC, the L2 controller of each core is built with prefetchers that can issue prefetches. We propose \emph{prefetch/request ratio} to quantify the prefetcher utilization. We use L3 APC (Accesses Per memory active Cycle)~\cite{wang2014apc} to denote off-core access performance. All the used metrics can be measured by performance counters on commodity processors.

\subsection{Pipeline Efficiency}
%In summary, we conclude that
\begin{itemize}
\item \emph{Most hotspots have low pipeline efficiency in terms of IPC.}
\item \emph{The hotspots in emerging workloads including Graph500, HPCC, and HPCG have the lowest IPC.}
\item \emph{A static instruction of a Bigdata application runs many times to process different data elements.}
\item \emph{Most hotspots in HPCC, HPCG, SPEC CPU2006 and PARSEC have very large pipeline stall degrees.}
\end{itemize}

We use three metrics to quantify the pipeline efficiency. Apart from IPC, the number of instructions executed per clock cycle,
we define a new metric, \emph{pipeline stall degree}. When the number of instructions and clock frequency are fixed, IPC indicates the final performance. To provide more details behind the IPC value, we define pipeline stall degree as the ratio of pipeline stall cycles to the total execution cycles, and define latency non-hidden degree as the ratio of pipeline stall cycles to the total memory active cycles.

\begin{figure}
	\centering
	\includegraphics[width=0.8\linewidth]{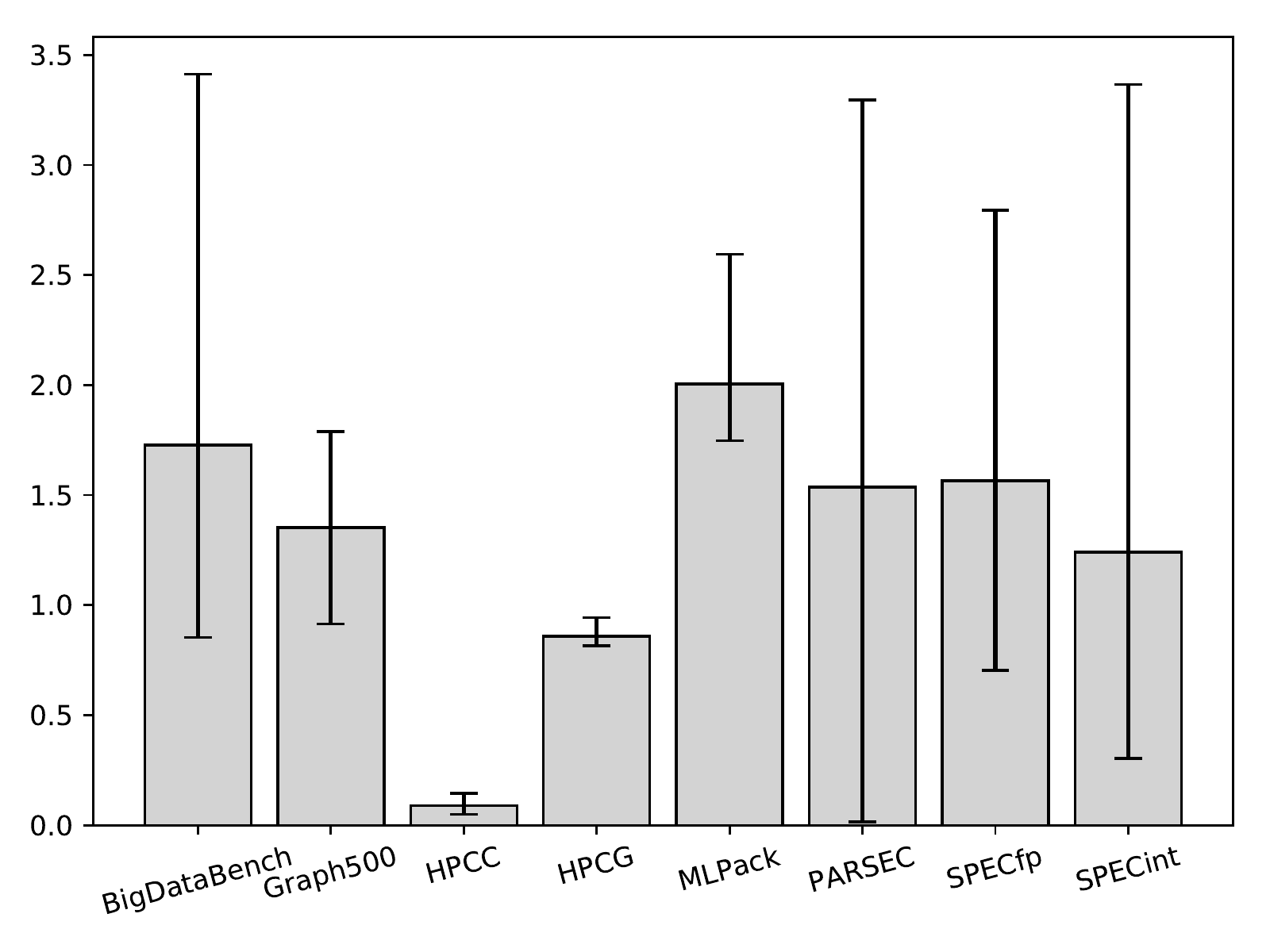}
	\caption{The IPC of the hotspots in diverse benchmark suites}
	\label{fig:IPC}
\end{figure}

Fig.~\ref{fig:IPC} shows the IPC of the hotspots. On average, the hotspots in Graph500, HPCC, and HPCG have low IPC, i.e., low efficiency. 
Therefore, we can infer that there exists a mismatch between the demands of these emerging applications 
and the micro-architecture of predominant modern processors. For instance, the data structures of Graph500 incur many indirect memory accesses which are challenging for current mainstream architectures.

\begin{figure}
	\centering
	\includegraphics[width=0.8\linewidth]{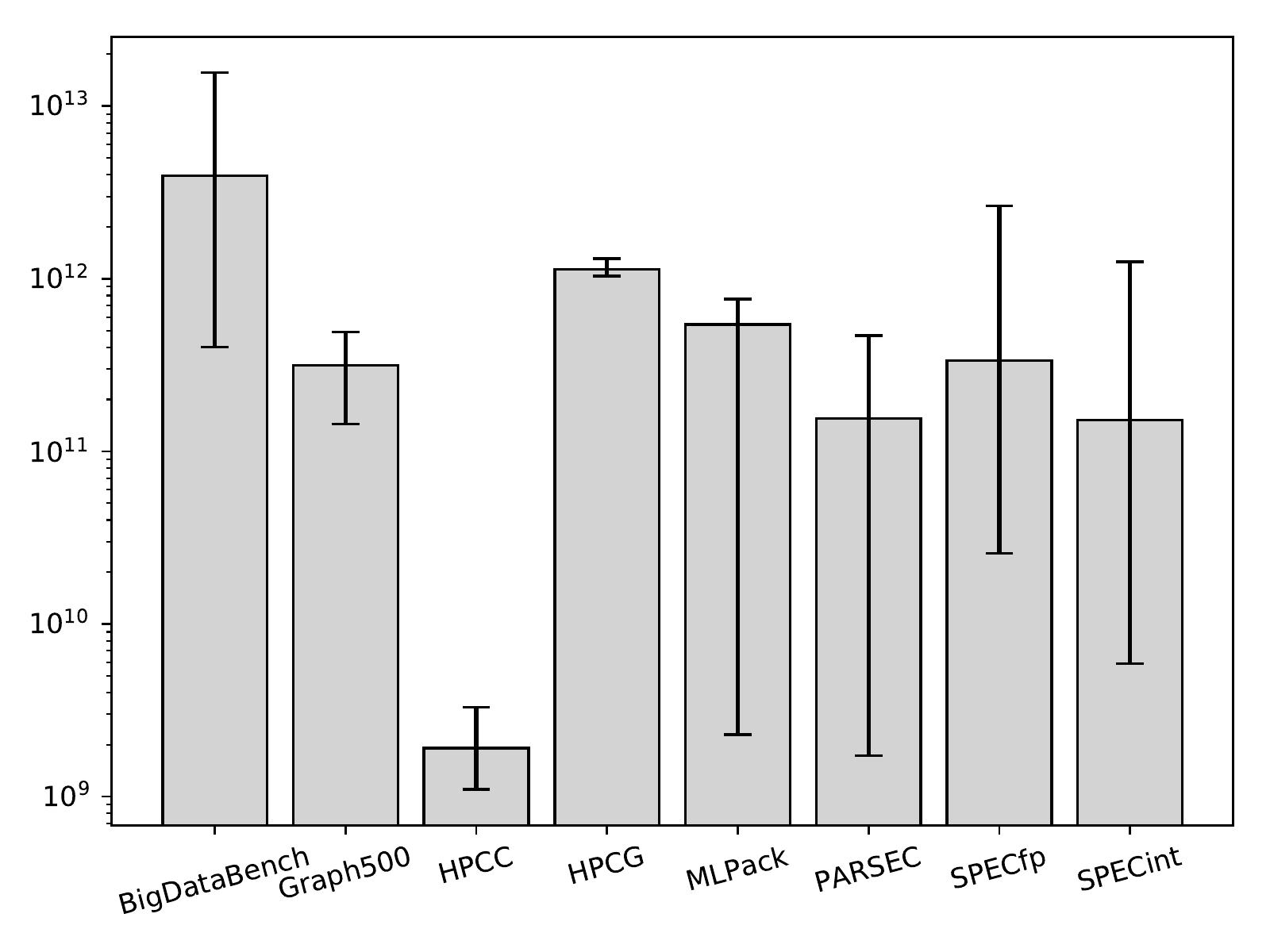}
	\caption{The dynamic instruction count (IC) of the hotspots in diverse benchmark suites}
	\label{fig:IC}
\end{figure}

\begin{figure}
	\centering
	\includegraphics[width=0.8\linewidth]{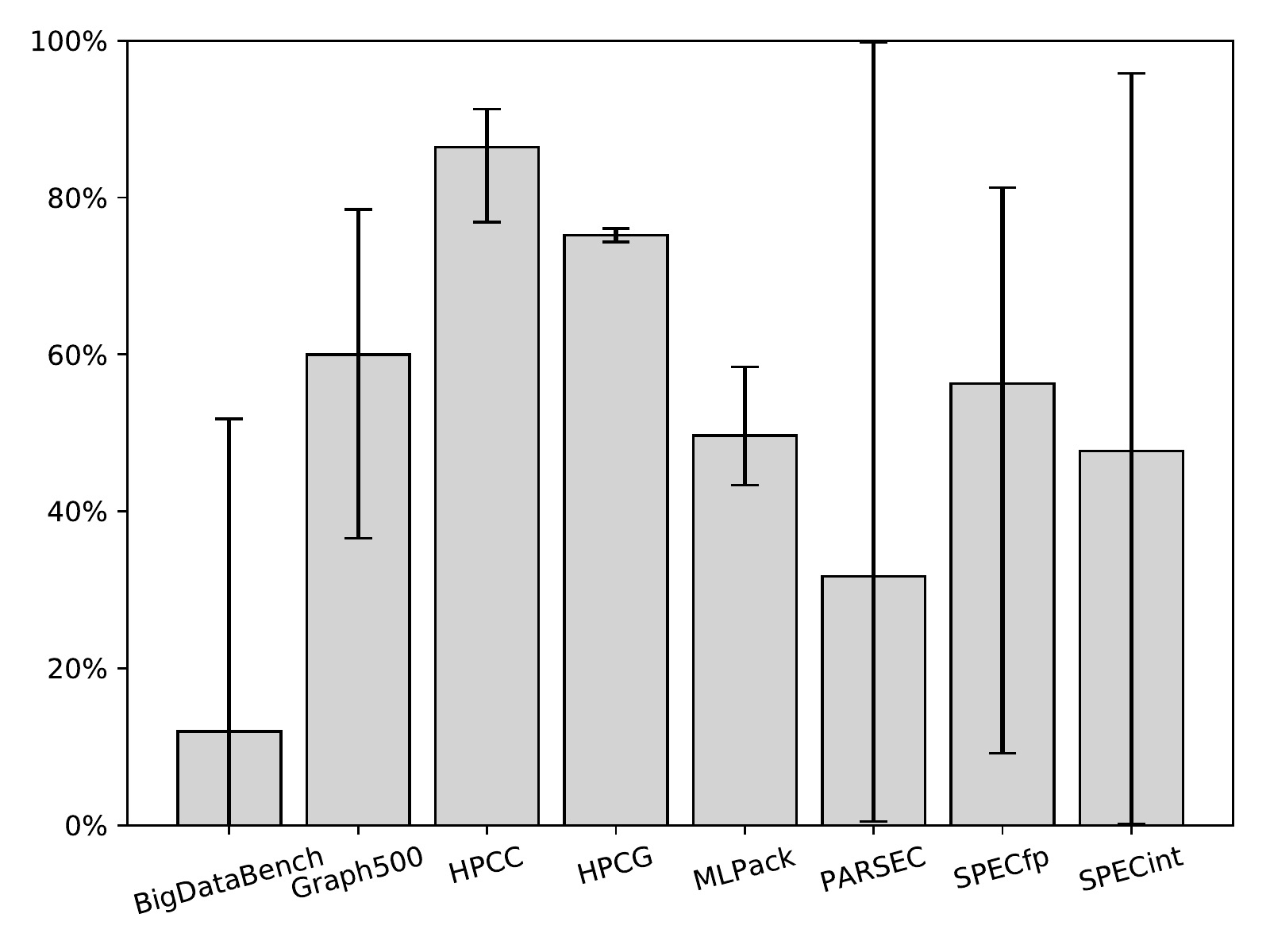}
	\caption{The pipeline stall degree of the hotspots in diverse benchmark suites}
	\label{fig:pipeline-stall-degree}
\end{figure}

Fig.~\ref{fig:IC} shows the workload size of the hotsopts in terms of the number of dynamic instructions (IC), while 
Fig.~\ref{fig:pipeline-stall-degree} presents the hotspots' pipeline stall degrees, 
%which is defined as the ratio of pipeline stall cycles over the total runing cycles.

We find that the number of static instructions of all the hotspots is in the same order of magnitude, 
which is defined by only a few lines of code. The hotspots in BigDataBench have executed the largest number of dynamic instructions. 
However, the pipeline stall degree of BigDataBench is the smallest. 
We can infer that a static instruction of BigDataBench has been run many times to process different data elements.

Pipeline stall directly results in the low efficiency of a computing system. 
On average, the hotspots in HPCC, HPCG, SPEC CPU2006 and PARSEC have very large pipeline stall degrees. 

\subsection{On-chip Cache Effectiveness}
\begin{itemize}
\item \emph{The hotspots in emerging applications, HPCC, HPCG and Graph500 have low on-chip cache effectiveness in terms of RaL~\cite{Liu2017CaL}, since a cache line has been reused few times before being evicted from on-chip caches.}
\item \emph{The differences among the on-chip cache effectiveness of different hotspots are quite significant; 
some can be more than 255 thousand times more effective than others.} 
\end{itemize}

Locality of data accesses is the fundamental principle that drives the hierarchical memory system design. Given the significance of the locality principle, previous works have attempted to quantify the locality for deep understanding of reference patterns and guiding compiler and architecture design to exploit program locality. For temporal locality, the histogram of reuse distances~\cite{zhong2009program} or LRU (least recently used) stack distances ~\cite{cade2009balancing} can be computed based on a sequential address trace. For spatial locality, however, there is a lack of consensus for such a quantitative measure and several ad-hoc metrics ~\cite{li2006cmp,jiang2010reuse} are proposed based on intuitive notions. It is not easy to represent locality with only a single score. Moreover, with the consideration of the ability of measurement in real platforms, we found that reuse distance like metrics cannot be measured directly via performance counters in commercial processors.

It is known that the memory stall time stems from the large access time ratio among the memory levels (i.e., the latency effect~\cite{wulf1995hitting}) and the limitation of the pin bandwidth (i.e. the bandwidth effect~\cite{Kagi1996Memory}~\cite{Rogers2009Scaling}). Prior work uses MR (Miss Rate) or MPKI (Misses Per Kilo Instructions) to characterize locality~\cite{Muralidhara2011Reducing}~\cite{Qureshi2007Adaptive}, which makes sense theoretically. However, commercial processors inculde multiple cache levels, so it is not convenient to use the metrics of different levels simultaneously and using only one or two of them would be misleading.

We use \emph{Reuse-aware Locality (RaL)}~\cite{Liu2017CaL}, which refers to the average amount of L1 cache hits that can be met by one off-chip data movement. The larger the value of RaL, the more reuses of the data, and the lower the requirement for off-chip bandwidth, since each off-chip data movement carries one cache line, the size of which is fixed in predominant commercial processors, i.e., 64B. With only one score value, RaL quantifies the effectiveness of the on-chip memory hierarchy. On-chip cache hierarchy takes most of the chip area and transistors. However, not all the types of applications can obtain the same large benefit as that in Bigdata applications.

\begin{figure}
	\centering
	\includegraphics[width=0.8\linewidth]{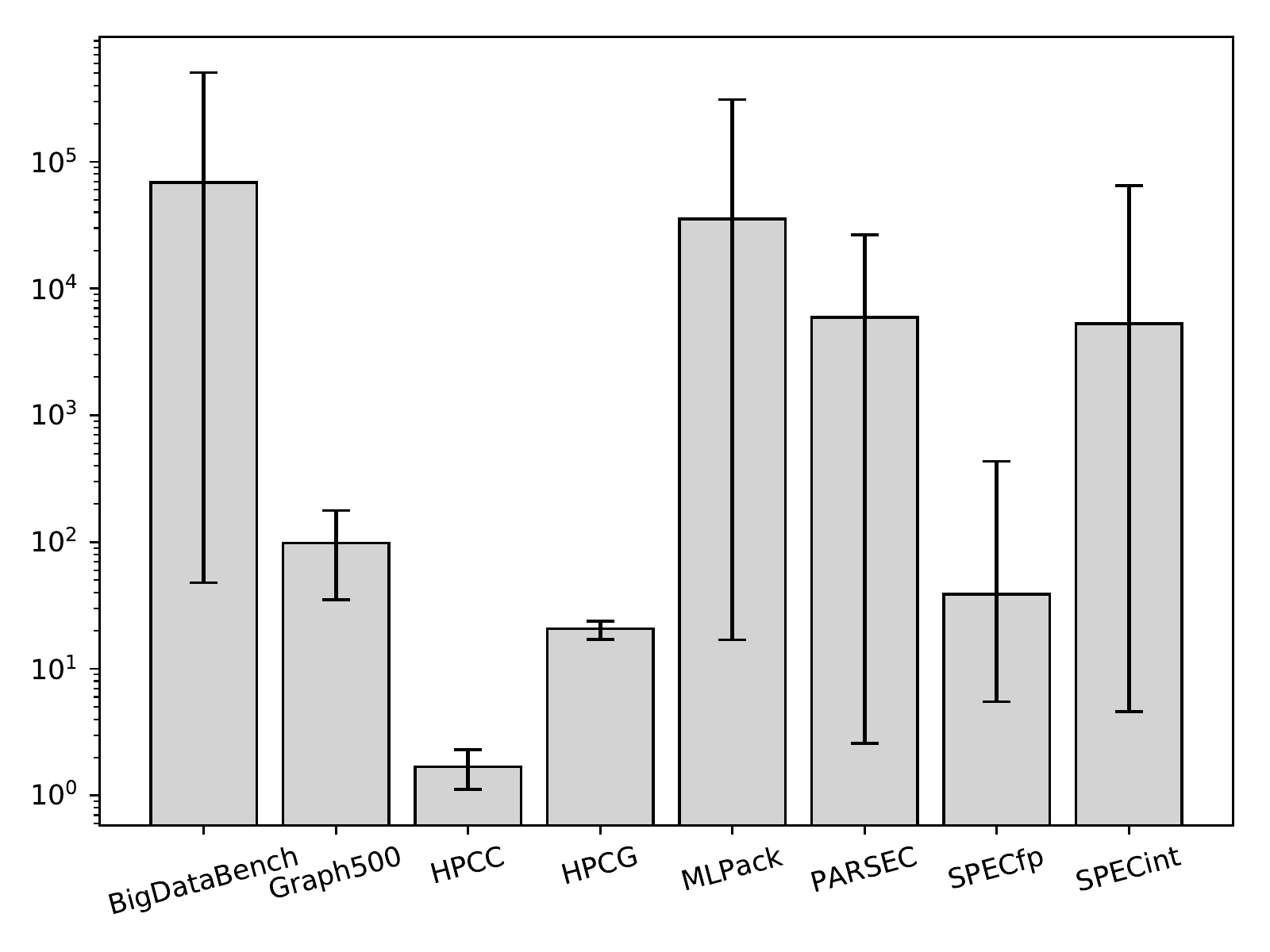}
	\caption{The Reuse aware Locality (RaL) of the hotspots in diverse benchmark suites}
	\label{fig:RaL}
\end{figure}

In Fig.~\ref{fig:RaL}, we can see that a significant portion of the Bigdata benchmarks have high RaL values. For these hotspots, once the data has been loaded into the on-chip cache hierarchy, they can be reused many times until being moved out of the processor chip. The hotspots in emerging applications, HPCC, HPCG and Graph500 have low RaLs.

A hotspot from $Random Access$ in HPCC suite has the least RaL of 1.98, while a hotspot from $wordcount$ in the BigDataBench has the largest RaL of 506,081. The difference is quite significant, with $wordcount$ being 255 thousand times more effective than $RandomAccess$. On the other hand, we found that 20 out of 34 hotspots of SPECint and SPECfp have small RaLs ($\textless$ 100). The small RaL implies that even for single-thread programs on-chip caches have not been efficiently utilized, i.e., many cache lines are dead blocks during most of their lifetime. This fact calls for a change in the trajectory of processors.

Besides the RaL, we define two additional metrics, $L2~ and~ beyond~ active~ degree$ and $latency~ non\text{-}hidden~ degree$. L2 and beyond active degree is the ratio of L1 pending cycles to total memory (including L1) active cycles, thus it shows how busy the memory system is. Latency non-hidden degree is the ratio of pipeline stall cycles to total memory (including L1) active cycles, thus it quantifies to what degree the memory access latency has not been hidden.

\begin{figure}
	\centering
	\includegraphics[width=0.8\linewidth]{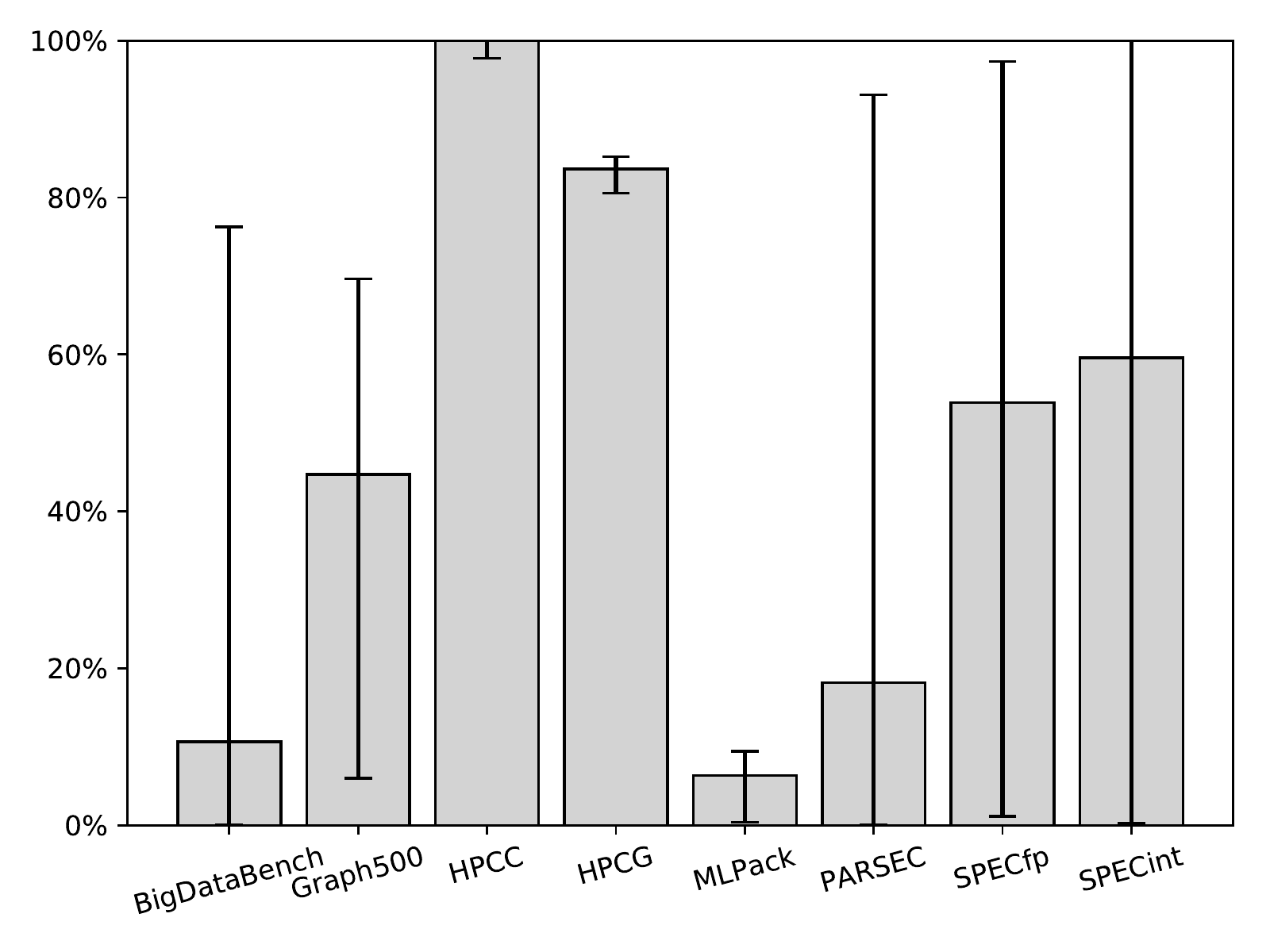}
	\caption{The L2 and beyond active degree of the hotspots in diverse benchmark suites}
	\label{fig:figure-L2-and-beyond-active-degree}
\end{figure}

As shown in Fig.~\ref{fig:figure-L2-and-beyond-active-degree}, on average the hotspots in HPCC and HPCG have large L2 and beyond active degrees, which implies that the memory hierarchy is active during most of the execution time, and thus causes large energy consumption. In comparison, from Fig.~\ref{fig:IPC}, it is seen that the IPC of HPCC and HPCG are very low. Therefore, we can infer that the busy activities in the memory hierarchy are very inefficient, which implies a significant mismatch between the patterns of HPCC and HPCG with the micro-architecture in modern commodity processors.

\begin{figure}
	\centering
	\includegraphics[width=0.8\linewidth]{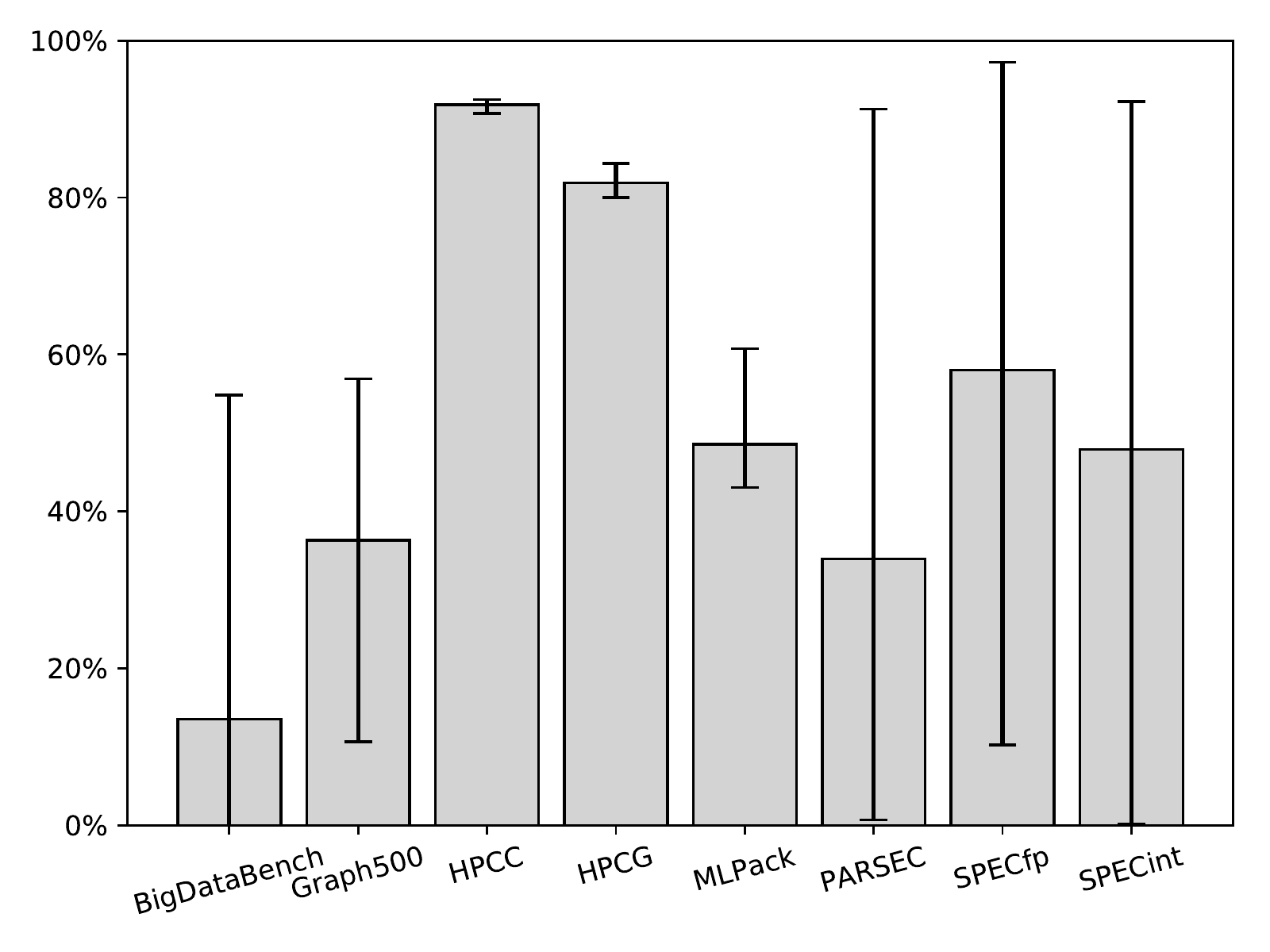}
	\caption{The latency non-hidden degree of the hotspots in diverse benchmark suites}
	\label{fig:figure-Latency-non-hidden-degree}
\end{figure}

As shown in Fig.~\ref{fig:figure-Latency-non-hidden-degree}, on average the hotspots in HPCC and HPCG have large latency non-hidden degree. The diversity of latency non-hidden degrees of hotspots in PARSEC, SPECint and SPECfp are high. The $mcf$ in SPECint, $milc$ in SPECfp, and $canneal$ in PARSEC have hotspots that are corresponding to the high latency non-hidden degrees.

\subsection{Prefetch/Request Ratio}

\begin{itemize}
\item \emph{The $RandomAccess$ in HPCC has the smallest prefetch percentage, while the HPCG has the highest.}
\item \emph{Prefetchers are effective for most hotspots except those in $RandomAccess$ from HPCC.}
\item \emph{Among the multithread benchmark suites except HPCC, Graph500's prefetcher has the lowest effectiveness.} 
\end{itemize}

\begin{figure}
	\centering
	\includegraphics[width=0.8\linewidth]{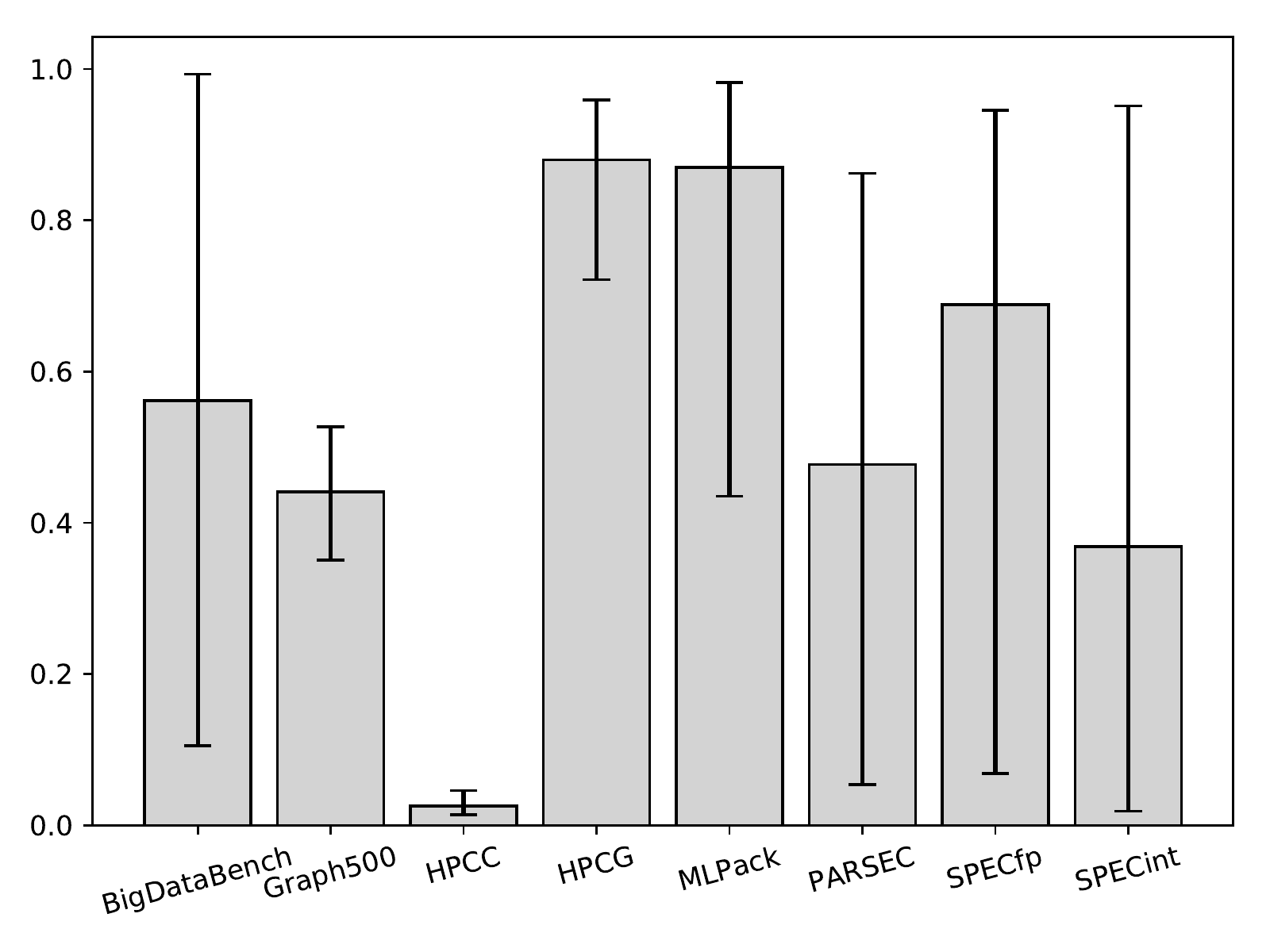}
	\caption{The L2 prefetch/request ratio of the hotspots in diverse benchmark suites}
	\label{fig:figure-L2_rqsts_ratio}
\end{figure}

In Intel Xeon processors~\cite{inteloptimize}, automatic hardware prefetch can bring cache lines into the unified last-level cache based on prior data misses. We find that there are two limitations of the Intel prefetcher. First, it attempts to prefetch only two cache lines ahead of the prefetch stream. Moreover, it requires some regularity in the data access patterns, i.e., a data access pattern has a constant and short stride. If the access stride is not constant, the automatic hardware prefetcher can partly mask memory latency if the strides of two successive cache misses are less than the trigger threshold distance.

The bar chart in Fig.~\ref{fig:figure-L2_rqsts_ratio} illustrates the ratio of L2 prefetches to L2 requests. The prefetches are triggered by L2 prefetchers according to the patterns of demand requests. On average, for most suites except for HPCC, prefetches occupy a large portion of requests. As shown in Fig.~\ref{fig:figure-L2_rqsts_ratio}, for each of the eight suites except HPCC, there exists at least one hotspot that has a high L2 prefetch/request ratio. The $RandomAccess$ in HPCC has the smallest prefetch percentage (2.6\% on average), while the HPCG has the highest prefetch percentage (88\% on average). According to the prefetch percentage, the hotspots are grouped into two categories, prefetch-friendly and prefetch non-friendly. We found that the prefetch-friendly hotspots have similar patterns that satisfy the trigger condition of Intel prefetch hardware. We visualize the patterns in Section 4. 

\subsection{Off-core Access Throughput}

\begin{itemize}
\item \emph{On average, single-thread applications have higher L3 APC than multithread applications.}
\item \emph{HPCC has the lowest L3 APC.}
\item \emph{Among the multithread benchmark suites, MLPack has the highest L3 APC}.
\end{itemize}

As RaL is an important leverage for memory performance optimization, once its before-optimization value is already high, to remove the hotspot, we should turn to other leverages such as increasing concurrency. However, once the RaL's before-optimization value is small, one can focus on improving RaL before turning to other leverages. 

To measure the off-core access performance, we have at least four choices: MLP, AMAT, Bandwidth, and APC. MLP refers to the memory level parallelism. AMAT is the average off-core access time. Bandwidth here means the off-core bandwidth consumption. 

APC~\cite{wang2014apc} refers to the number of accesses completed per memory active cycle, and is a measurement that considers both AMAT and MLP~\cite{chou2004microarchitecture}. Thus, APC model is comprehensive and reflects the quality of service (QoS) of memory system. Specifically, no matter how long cache hit latency is or how high cache hit rate is, the concern is the final service quality, i.e., how many accesses are finished in each memory active cycle. Due to these merits, we use APC in our study.

\begin{figure}
	\centering
	\includegraphics[width=0.8\linewidth]{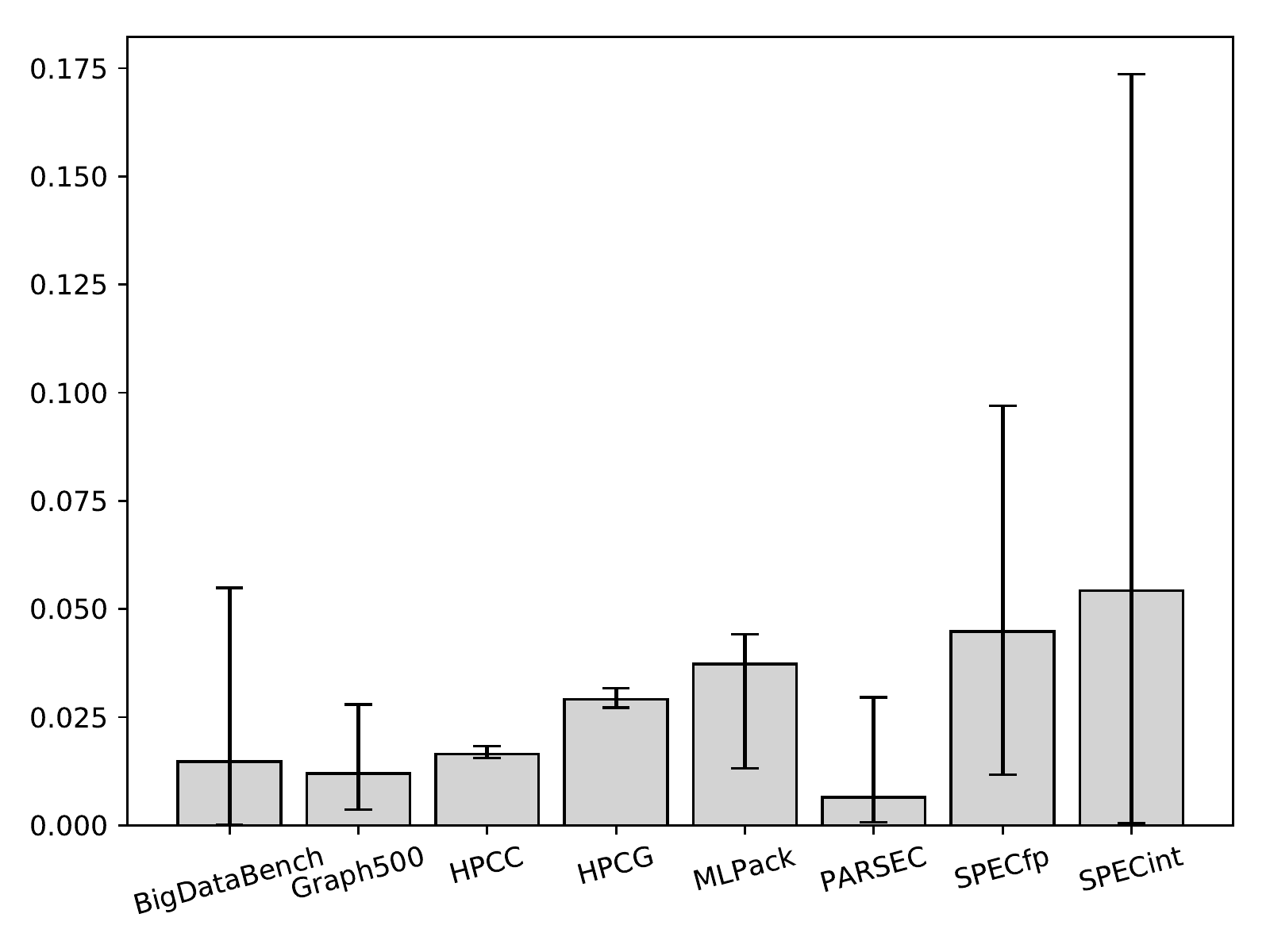}
	\caption{The L3 Accesses Per memory active Cycle (APC) of the hotspots in diverse benchmark suites}
	\label{figure-l3-apc}
\end{figure}

Fig.~\ref{figure-l3-apc} shows the APC value of L3 cache, which reflects how quickly the off-core accesses can be completed by L3. The hotspots in PASRSEC, Graph500, HPCC and HPCG have low L3 APC, which implies low off-core access throughput. 

\begin{figure}
	\centering
	\includegraphics[width=0.9\linewidth]{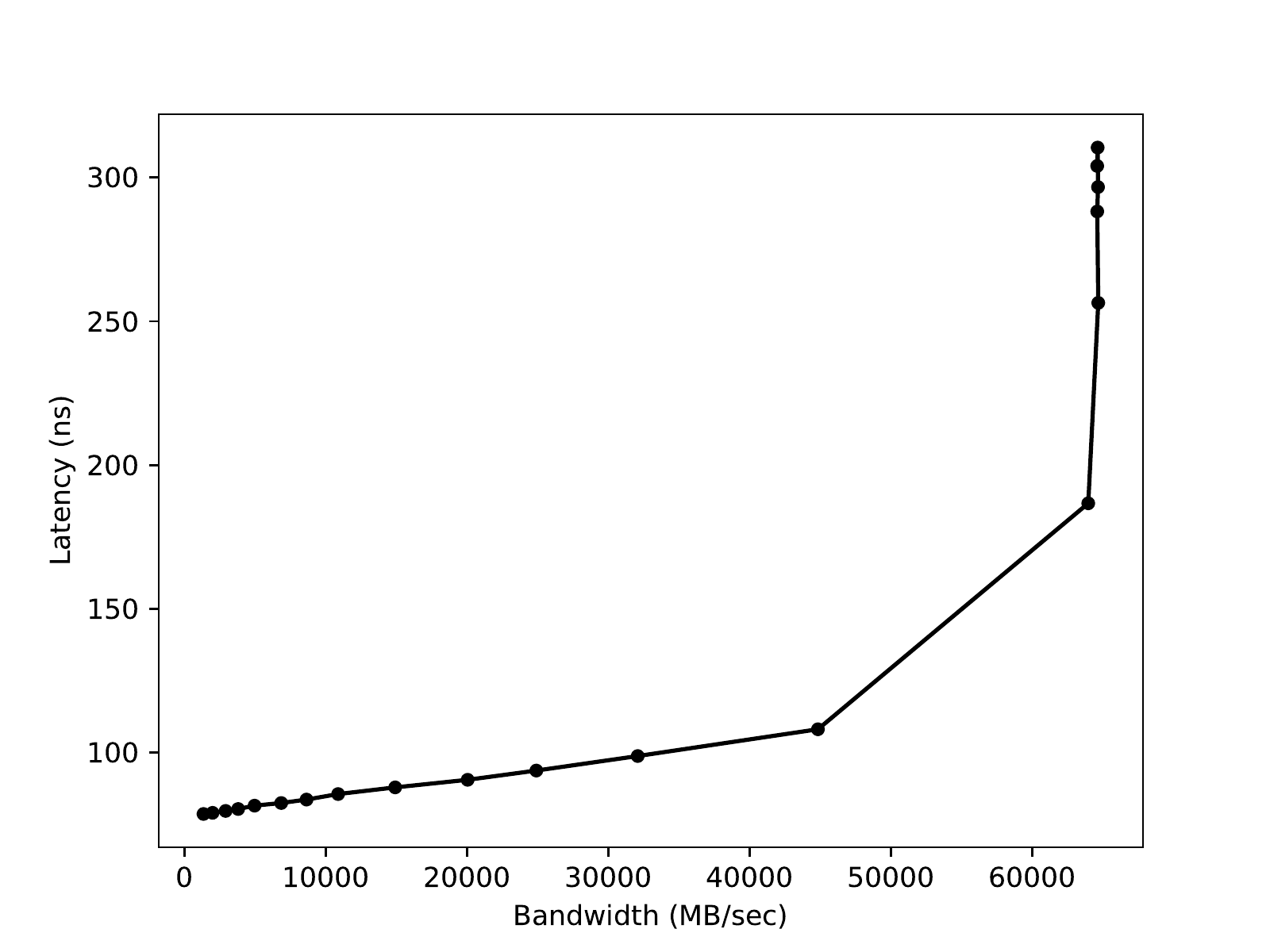}
	\caption{The memory access latency increases with bandwidth}
	\label{fig:figure-latency}
\end{figure}

By Little's law, L3 APC equals the value of MLP divided by memory access latency. Given the same access intensity of each thread, the more concurrent threads, the higher the bandwidth utilization. We measure memory access latency using Intel MLC (memory latency checker)~\cite{MLCIntel}. As shown in Fig.~\ref{fig:figure-latency}, the memory access latency increases with the bandwidth utilization. The hotspots in SPECint and SPECfp have high L3 APC, implying that single-thread applications have shorter contention delay than multithread applications. Thus, it calls for enhancing NoC and memory controllers of CMP for boosting the L3 APC of multithread applications. 

\section{Gene-Patterns and Periodic Table}
\label{sec:vp}

\begin{itemize}
\item \emph{(P$_1$,...,P$_6$) is a minimal complete set of base patterns, which is taken as Gene-Patterns.}
\item \emph{The Periodic Table of patterns can be built based on RaL and L3 APC, and is similar to the energy level diagram in physics, where the memory performance optimization is a transition between two energy levels.}
\item \emph{For the average pipeline stall degree, HPCC > HPCG > Graph500 > SPECfp > MLPack > SPECint > PARSEC > BigDataBench as evidenced by the order of the indifference curves in the Periodic Table.}
\item \emph{High L3 APC and high RaL rarely occur simultaneously for the hotspots of the benchmarks from diverse domains.}
\item \emph{The location in Periodic Table quantifies the matching degree between a pattern and an architecture.}
\end{itemize}

\subsection{Visualizing the Patterns and Finding the Gene-Patterns}
\label{sec:vp-gene}

\begin{figure}
	\centering
	\includegraphics[width=0.8\linewidth]{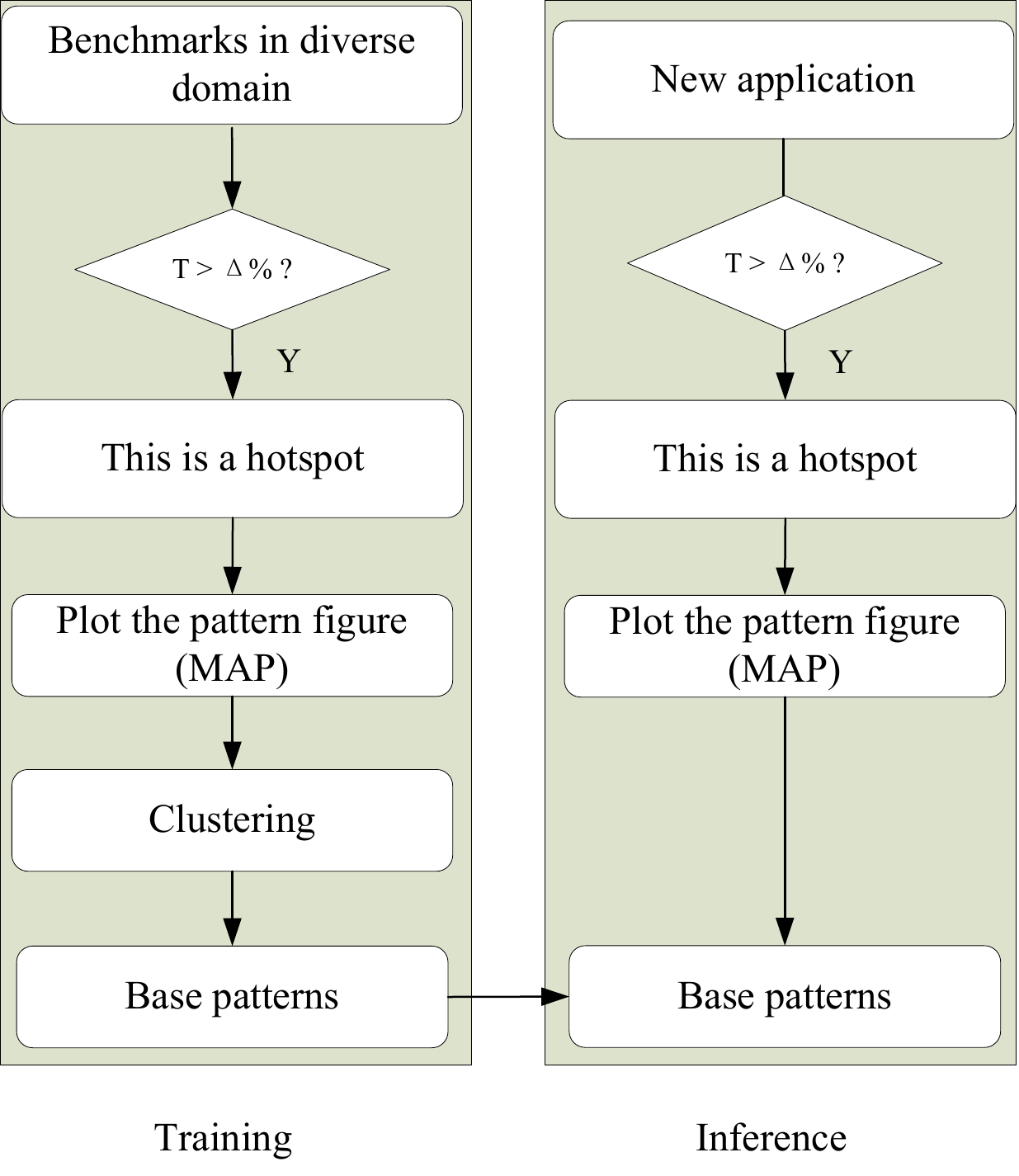}
	\caption{The two phases of pattern analysis for any application: training and inference.}
	\label{fig:2}
\end{figure}

The flow chart in Fig.~\ref{fig:2} illustrates how we conduct the recursive reduce process. We run the benchmarks and identify their hotspots. Each hotspot has a few pattern figures. We have collected 2420 pattern figures, which constitute the library of patterns. Although the size of the library is very large, we find that there exist significant similarities between the patterns of hotspots. Based on the similarities, the base patterns are extracted from the pattern library by clustering. 

\begin{figure*}
	\centering
	\includegraphics[width=0.92\linewidth]{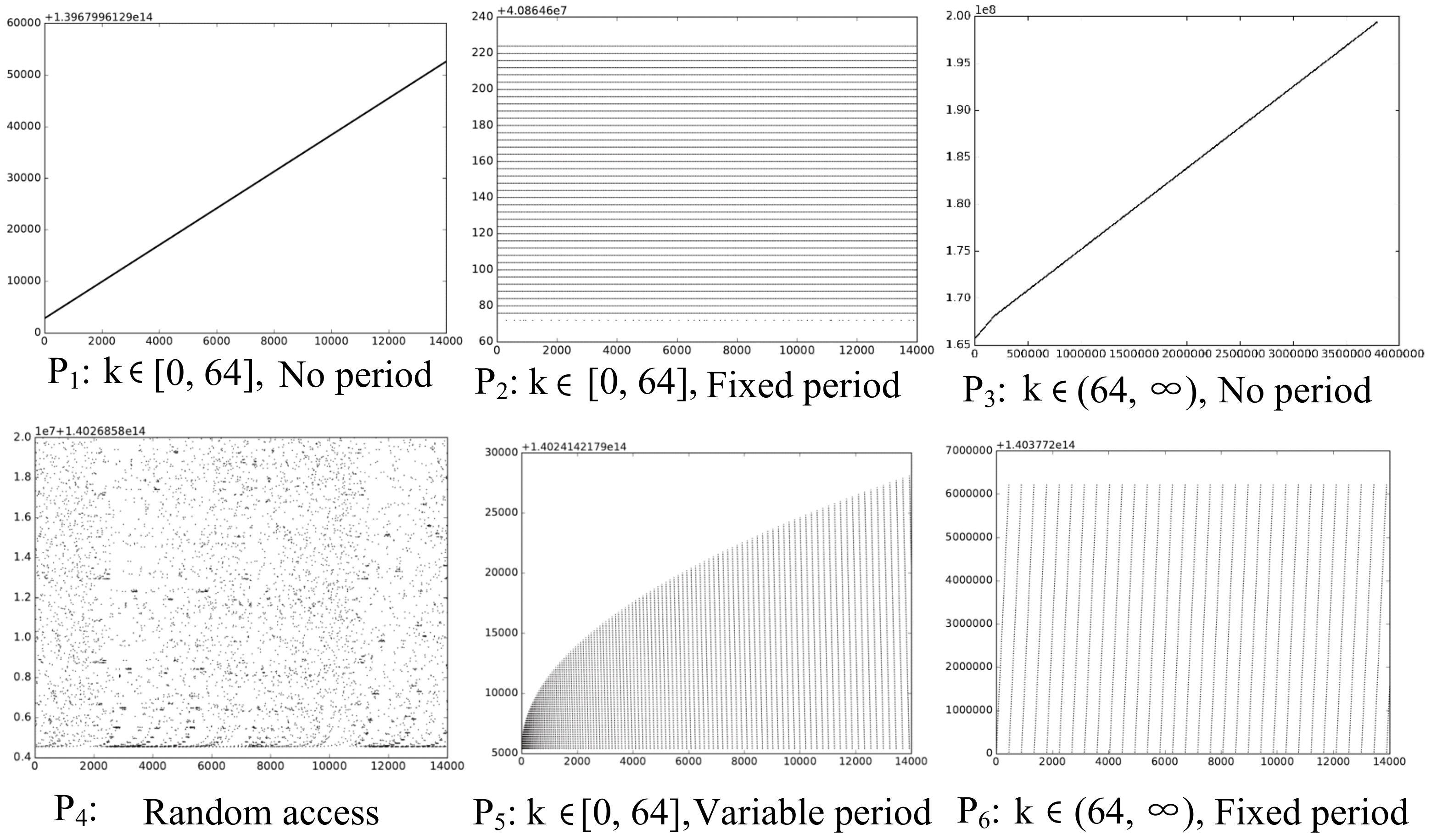}
	\caption{A minimal complete set of base patterns for the huge pattern space. Each base pattern figure is extracted from the pattern library.}
	\label{fig:3}
\end{figure*}

Following the workflow in Fig.~\ref{fig:2}, we find six different base patterns, which account for only 0.25\% (6/2420) of the pattern library. We plot the base patterns in Fig.~\ref{fig:3}, where the horizontal axis specifies the sequence of the accesses, and the vertical axis indicates the accessed logical address range. 

Pattern P$_1$ and P$_3$ are straight lines and aperiodic, the slope of which indicates how quickly the touched address space is extended. The larger the slope ($k$), the larger the working set. Specifically, when the slope ($k$) is zero, the data can be reused for each access by following the first access. On the other hand, when the slope ($k$) is infinite, there is no opportunity for data reuse.

Pattern P$_2$, P$_5$ and P$_6$ are straight lines and periodic. Compared to pattern P$_1$ and P$_3$, pattern P$_2$ and P$_6$ have more opportunities for data reuse due to their periods. Compared to P$_2$ and P$_6$, pattern P$_5$ has a variable period. The variability of the period impacts the reuse opportunities.

The straight line pattern figures are friendly for prefetchers. Pattern P$_4$ has no straight line but randomly distributed points, which would result in the case that the prefetchers cannot work well.

Providing the minimal complete set (MCS) of patterns is of great significance for architecture and software design. When pattern space is huge, there could exist multiple MCSs. We can prove that {P$_1$, P$_2$, ..., P$_6$} is a MCS.

\emph{\textbf{Theorem-1: $($P$_1$, P$_2$, ..., P$_6$$)$ is a minimal complete set of data access patterns.}}

$Simplified$~\begin{prf}$:$ 
We can prove (omit here) that any subset of (P$_1$, P$_2$, ..., P$_6$) is not a complete set. Then we only need to prove that (P$_1$, P$_2$, ..., P$_6$) is a partition of pattern space $S$. First, we have that any two different patterns are exclusive, i.e.,

\begin{align}
\label{exclusive}
P_i \cap P_j = \emptyset~~( 1 ~\le ~i, ~j \le ~6, ~i ~\neq ~j)
\end{align}

Second, the union of P$_1$ and P$_3$, P$_1$ $\cup$ P$_3$, represents all the patterns whose pattern figures are straight lines and aperiodic, while the union of P$_2$ and P$_6$, P$_2$ $\cup$ P$_6$, represents all the patterns whose figures are straight lines and have fixed periods. P$_5$ represents the patterns, the figures of which are straight lines with low slope and have variable periods. For the patterns the figures of which are straight lines with high slope and have variable periods, they can be taken as a special case of Random Access pattern, P$_4$. For patterns whose pattern figures are not straight lines, they can be classified into Random Access pattern, P$_4$. Therefore, we obtain that $\bigcup_{k=1}^{6}P_k=S$. 

By combining the above two aspects, we conclude that (P$_1$, P$_2$, ..., P$_6$) is a partition of pattern space $S$. \qed
\end{prf} 

\begin{figure*}
	\centering
	\includegraphics[width=0.86\linewidth]{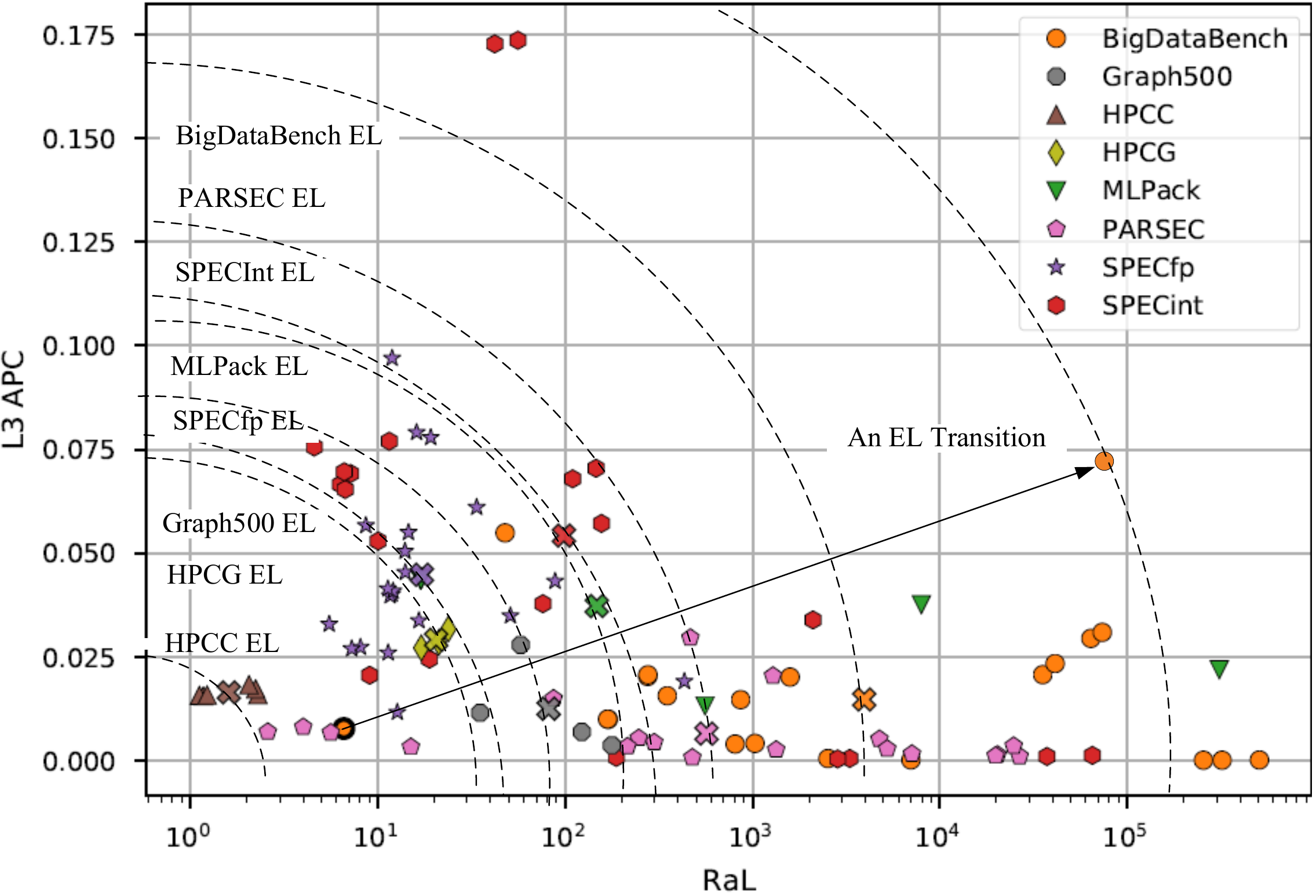}
	\caption{The periodic table of the hotspots in diverse benchmark suites. The dash lines are the Indifference Curves, which are analogous to energy levels (EL) in physics. Each cross marks the geometrical center of a benchmark suite. The EL transition is due to an optimization for $K\text{-}Means$ which will be introduced in Section 6.}
	\label{fig:figure-ral-l3apc}
\end{figure*}

\subsection{Periodic Table of Memory Access Patterns}
\label{sec:pt-map}

Depending on the matching degree between the microarchitecture and the Gene-Patterns, the memory access performance of hotspots is different.

In our experiments, among many metrics, the combination of RaL and L3 APC is found to be most effective. Fig.~\ref{fig:figure-ral-l3apc} shows both the RaL and L3 APC of each hotspot. The upper-right corner is blank, implying that high L3 APC and high RaL rarely occur simultaneously for the hotspots of benchmarks from diverse domains. The hotspots in HPCC and HPCG are close to the lower-left corner, which implies why they have large latency non-hidden degree and pipeline stall degree. Fig.~\ref{fig:figure-ral-l3apc} is the Periodic Table of Memory Access Patterns (PT-MAP), which can explain the results of latency non-hidden degree in Fig.~\ref{fig:figure-Latency-non-hidden-degree} and pipeline stall degree in Fig.~\ref{fig:pipeline-stall-degree}.

For memory access performance, the on-chip effectiveness (RaL) can substitute off-core memory access concurrency (L3 APC) and vice versa. An indifference curve is a period, which depicts that the RaL is substitutable for L3 APC. Fig.~\ref{fig:figure-ral-l3apc} shows eight indifference curves. Each curve corresponds to a benchmark suite, since the average value of RaL and L3 APC of the benchmark suite (marked as a cross) is on the curve.

Following the indifference curves in Fig.~\ref{fig:figure-ral-l3apc}, we find that the eight benchmark suites can be ordered according to the location of their indifference curves from the bottom left to the upper right: HPCC, HPCG, Graph500, SPECfp, MLPack, SPECint, PARSEC, and BigDataBench. This order is in agreement with the average pipeline stall degree from large to small shown in Fig.~\ref{fig:pipeline-stall-degree}. The agreement demonstrates that the location in the PT-MAP quantifies the matching degree between a pattern and an architecture. 

The PT-MAP is similar to the energy level diagram in physics, where a memory performance optimization essentially is a transition between two energy levels. In Fig.~\ref{fig:figure-ral-l3apc}, as the dash lines move from the lower left quarter to the upper right quarter, it transits to increasingly higher energy levels where the memory bound effect is lower. Of practical significance, in PT-MAP, both L3 APC and RaL can be measured in commodity processors. Such a simple, quantitative, and effective periodic table is timely and useful as applications are increasingly data intensive and diverse.

To enable an energy level transition, we can conduct Gene-Pattern aware accelerator design and software development.

\section{Gene-Pattern aware Accelerators}
\label{sec:bg}

In mathematics, the linear space has a series of basic vectors called ``bases'', and any vector in a linear space can be represented by a linear combination of the bases. If we think of applications as ``vectors'' in a linear space, then all we need to do is find the corresponding ``bases''. Once we find the ``bases'' of applications, we need only focus on the bases, the types of which are very limited. In that sense, the computing system design process can benefit from ``linear algebra''.

The base of the pattern space $V$ is (P$_1$,P$_2$,...,P$_6$). For any pattern $P$ in $V$, $P=\alpha_1 \times P_1 + \alpha_2 \times P_2 + ... +\alpha_n \times P_n$, here ``+'' is ``$\cup$''. Table~\ref{tab:table2} shows the pattern distribution of different benchmark suites. When a hotspot has a mixed pattern, we use weights in the statistical analysis. Table~\ref{tab:table2} motivates us to build Gene-Pattern aware accelerators.
 
Both pattern P$_1$ and P$_3$ can use prefetching, since their pattern figures are straight lines. However, they also have significant differences. For pattern P$_1$, the slope of the straight line is small, so the spatial locality is high. Therefore, for pattern P$_1$, open page policy of DRAM can be used, which brings more benefits than closed page policy; the cache line size can be coarse-grained for effective prefetching. 

On the other hand, for pattern P$_3$, the slope of the straight line is large, so the spatial locality is low and the working set size increases quickly. As a result, for pattern P$_3$, close page policy of DRAM brings more benefits than open page policy; the fine-grained cache line size rather than coarse-grained will reduce bandwidth consumption and mitigate contention, thus improving performance and reducing power consumption.

Pattern P$_4$ is characterized by Random Access and has low spatial locality, so it is preferable to use fine fetching granularity and close-page policy. Pattern P$_4$ has a limited degree of temporal locality. The same data is read and then is written immediately, and will rarely be reused in future. Therefore, for Pattern P$_4$, we only need to cache the corresponding data in L1 rather than L2 and L3. In fact, if we allocate space for the data of pattern P$_4$ in L2 and L3, the working set increases exponentially and thus consumes the limited cache space quickly, which would flush the data of peer programs and severely impact their performance.

For Pattern P$_6$, the slope of the straight line is very large but the period is fixed. We can use fine-grained prefetch and fetch. Because the slope is very large, the working set increases sharply, so caching the data in L1 and L2 is of no use for performance. The data can be allocated in LLC, especially in die-stacked DRAM cache that is several GBs. Similar to P$_3$, P$_6$ prefers the close-page policy.

Both pattern P$_2$ and P$_5$ have low slope, but their periods are different, one is fixed, and the other is variable. They both can use caching and coarse-grained fetch. However, their prefetch policies are different.

Following the above hints, we can build hardware accelerators for each of the Gene-Patterns. As the ``genes'' of applications have already been found, highly customized accelerators for each ``gene'' can be designed. When an application is executing, hardware can detect the ``genes'' of the application. Then, the accelerators of the used genes are enabled and the accelerators of the unused genes are disabled. In this manner, the computer architecture adapts to the software diversity automatically to harvest both performance and energy-efficiency.

%\begin{scriptsize}
\begin{table}[h!]
  %\scriptsize
\centering
\caption{The pattern distribution of diverse benchmark suites}
\label{tab:table2}
\begin{tabular}{|c|c|c|c|c|c|c| }
\hline  % \\ 
\emph{Suites} & $P_1$ & $P_2$ & $P_3$ & $P_4$ & $P_5$ & $P_6$\\ [0.5ex] 
\hline
\emph{HPCC} & 0 & 0.25 & 0 & 0.75 & 0 & 0 \\
\hline
\emph{HPCG} & 0 & 0.25 & 0.75 & 0 & 0 & 0 \\
\hline
\emph{Graph500} & 1 & 0 & 0 & 0 & 0 & 0 \\
\hline
\emph{SPECfp} & 0.47 & 0.42 & 0.02 & 0 & 0.09 & 0 \\ 
\hline
\emph{MLPack} & 0 & 0.43 & 0 & 0 & 0 & 0.57 \\ 
\hline
\emph{SPECint} & 0.04 & 0.56 & 0 & 0.03 & 0.11 & 0.26 \\ 
\hline
\emph{PARSEC} & 0.17 & 0.64 & 0 & 0.18 & 0 & 0 \\ 
\hline
\emph{BigDataBench} & 0.14 & 0.81 & 0.02 & 0.03 & 0 & 0 \\
\hline
\end{tabular}
\label{table:Table} 
\end{table}
%\end{scriptsize}

\section{Gene-Pattern aware Programming}
\label{sec:rg}

\begin{itemize}
\item \emph{Dynamic data structures can change the data reuse period and the working set.}
\item \emph{Abuse of pointers is harmful for memory access performance, similar to the ``go to'' statement.}
\item \emph{The compound data structures usually have low memory access locality.}
\item \emph{Accessing array in column rather than row order would lower the access locality.}
\item \emph{Random access pattern is responsible for the inefficiency of many hotspots.}
\end{itemize}

As with hardware design, the development of software including compilers and applications can also be inspired by the Gene-Patterns, since the Gene-Patterns have deepened our understanding about the impact of data structure and algorithm on memory access patterns.

Table~\ref{tab:table3} shows the abstract code example of the base patterns. Data structures indicate how the data is represented. There are three fundamental types of static data structures: $array$, $record$, and $set$. They constitute the building blocks out of which more complex structures are formed. For the dynamic structures, not only the values but also the structures of variables are changed during the computation. Dynamic structures include $linked~list$, $tree$, and $graph$, where pointers play an important role in accessing the data elements. 

Dynamic data structures can change the data reuse period and the working set of the memory accesses. For instance, a hotspot in $mcf$ from SPECint has pattern P$_5$. The hotspot traverses a linked list and inserts new elements into the list in each iteration of the loop, so the linked list is gradually increasing, which results in pattern P$_5$.

The pointers are flexible for programming but hurt data access locality. We find another hotspot of $mcf$ is traversing a heap, and the pattern is a combination of P$_4$ and P$_5$. $Heap$ is a special case of a complete binary tree. When traversing a branch of the complete binary tree, the stride between two continuous accesses increases exponentially, which results in pattern P$_4$. The addresses having been accessed tend to increase over time, which indicates that there is an upward trend of the amount of the elements in the heap. That trend results in the Pattern P$_5$.

We found that the compound data structures usually have low memory access locality. For instance, when an array's index is an element of another array, the memory access would have low locality. HPCG has pattern P$_3$, where the corresponding code is Loop {A[B[j]]}. The A[B[j]] is an indirect access of array A, since the index of A is the element of array B.

%\begin{scriptsize}
\begin{table}[h!]
  %\scriptsize
%\centering
\caption{The abstract code examples of the base patterns}
\label{tab:table3}
\begin{tabular}{|c|c|c| }
\hline  % \\ 
\emph{Patterns} & \emph{Example Code} & \emph{Note} \\ [0.5ex] 
\hline
$P_1$ & \tabincell{c}{Loop i \\\{ Operate A[i]\}} & stride of A[i] > $c$ \\
\hline
$P_2$ & \tabincell{c}{Loop i \{\\ Loop j \{ \\ Operate A[j] \} \}} & stride of A[j] < $d$ \\
\hline
$P_3$ & \tabincell{c}{Loop i \\\{ Operate A[B[i]] \}} & stride of B[i] > $d$ \\
\hline
$P_4$ & \tabincell{c}{Loop i \\\{ Operate A[random()]\}} & A is large \\ 
\hline
$P_5$ & \tabincell{c}{For i = 1 to n \{ \\ For j = 1 to i \{ \\ Operate A[j] \}\}} & stride of A[j] > $c$ \\ 
\hline
$P_6$ & \tabincell{c}{Loop i \{\\ Loop j \{ \\ Operate A[B[j]] \} \}} & stride of B[j] > $d$ \\ 
\hline
\end{tabular}
\label{table:Table} 
*$c$ is the size of cache line (64B)~~~~~~~~~~~~~~~~~~~~~~~~~~~~~~~~~~~~~~~~~~~~~~~~\\
*$d$ is the trigger threshold distance of prefetch hardware
\end{table}
%\end{scriptsize}

We found that accessing array in column rather than row order would lower the access locality. As for the data structures whose implementations are based on array, the stride between two continuous accesses significantly influences to the final pattern. Different strides indicate different values of the slope, $k$. Accessing arrays in column order usually occurs for the hotspots in Graph500, which results in pattern P$_6$. Since the stride is large, $k$ is also large.

When random numbers are used as the index of array, pattern P$_4$ occurs, which has the lowest locality among all the patterns. We found that not only HPCC has random access pattern, but also that $canneal$ from PARSEC has random access pattern. In $canneal$, a continuous memory region instead of many small memory regions is allocated. Thus, although the memory access pattern is random access (pattern P$_4$), it still concentrates within a range of about 20MB.

Although the focus of our study is not on optimizing a special application, we conducted software optimizations for many benchmarks following the above analysis of Gene-Patterns. For instance, we found that the $importance$ of a hotspot in K-means from BigDataBench is 71.37\%. After our optimization with prefetching and splitting, the location of the hotspot in PT-MAP moves to the upper right corner. The energy level transition is shown in Fig.~\ref{fig:figure-ral-l3apc}. The total running time of the benchmark is reduced from 523.99s to 336.98s. That is, with a small number of code modifications, the total application performance has been improved by 55\%.

\section{Related Work}
\label{sec:rw}

Ganesh et al.~\cite{Venkatesh2010Conservation} proposed Conservation Cores (\emph{c-cores}), which are specialized processors that focus on reducing energy instead of increasing performance. The focus on energy makes \emph{c-cores} an excellent match for many applications that would be poor candidates for hardware acceleration (e.g., irregular integer codes). In contrast, our study is not only for reducing energy, but also for improving performance, because even the codes with irregular patterns can be accelerated.

In the work of PuDianNao~\cite{conf/asplos/LiuCLZZTFZC15}, one could observe that the variables in k-NN distance calculations are naturally clustered into three classes according to the average reuse distances (similar behaviors can be observed from K-Means). Therefore, PuDianNao uses three separate on-chip buffers in its accelerator, where each buffer stores the variables having similar reuse distances. Our results corroborate these findings, and more importantly, reveal the reasons behind the observations from the Gene-Pattern perspective.

Simulation is an alternative way to conduct the characterization~\cite{Barrow2009A}. In contrast, we run the applications in real machines and use performance counters to obtain accurate metrics. It is difficult to use software simulation methods to run each benchmark completely, especially those with large footprints. The speed of cycle accurate simulators such as GEM5 is pretty low~\cite{binkert2011gem5}, which is a typical slowdown of the real execution time in the orders of 10$^5$ to 10$^6$. For instance, we have measured the speed of GEM5, which is 50 to 500 KIPS (Kilo Instructions Per Second), much lower than the peak speed of Intel Xeon processors (~8 GIPS). Even simulating a relatively small program that takes one minute to execute requires approximately one month to a year to simulate~\cite{McKee2006}. Our proposed HOTS will benefit architects to conduct simulation with representative and concise inputs. 

As opposed to the previous research which tends to characterize only a special benchmark suite~\cite{Bienia2008The}~\cite{Wang2014BigDataBench}, our study broadly analyzes diverse applications from eight different suites, and compares them in depth. Our focus is not on a special benchmark suite, but the base patterns of all the different suites, which are the essential components of applications. To the best of our knowledge, our study is the first work that covers such a wide range of benchmark suites and extracts the base patterns of applications. 

Some existing research measures the average value of metrics on real machines to characterize the average behavior of applications~\cite{Bienia2008The}~\cite{Wang2014BigDataBench}. For example, CloudSuite ~\cite{Ferdman2012Clearing} performs a 3-minute measurement after a warmup phase. Although the average is a single value that is easy to present, the diversity of the patterns in an application is hidden by the average. In our work, we identify the application hotspots, which have a small number of source lines but consume significant running time. Running a whole application completely but characterizing only in the granularity of hotspot, renders our work different from most other studies.

Unlike Cloudsuite~\cite{Ferdman2012Clearing} and BigDataBench~\cite{Wang2014BigDataBench}, we propose an evalution method using Reuse aware Locality (RaL)~\cite{Liu2017CaL}, pipeline stall degree, L2 and beyond active degree, prefetch/request ratio and latency non-hidden degree. These metrics present new perspectives for characterizing the utilization of the micro-architectures. 

The work in BigDataBench~\cite{Wang2014BigDataBench} finds that L3 caches of a typical state-of-practice processor (Intel Xeon E5645) are efficient for big data workloads. Our results not only corroborate this finding, but also reveal other things with respect to the metric, RaL. The on-chip cache hierarchy is highly effective for most hotspots in BigDataBench, MLPack, PARSEC, SPECint, and SPECfp. However, for some emerging workloads, like Graph500, HPCC, HPCG, the on-chip cache effectiveness is very low.

The RaL metric~\cite{Liu2017CaL} we used is similar to the Roofline model~\cite{Williams2009Roofline}. In Roofline model, operational (arithmetic) intensity is the number of floating-point operations per byte of memory accesses. Roofline demonstrates that, if an application is bounded by the peak computing speed, it is compute-bound, while if it is bounded by the product of peak memory bandwidth and operational intensity, it is memory-bound. Compared to operational intensity, the RaL metric is the number of L1 cache hits per off-chip memory access, which is more focused on memory system performance. 

The substitution effect between cache size and memory bandwidth has been discussed in the REF work~\cite{Zahedi2014REF}. Our PT-MAP is similar to the Cobb-Douglas Indifference Curves of REF. In contrast, PT-MAP is closer to the essence of memory access patterns. We replace cache size with RaL, and prefer L3 APC to memory bandwidth. PT-MAP is simple and effective, and the energy level concept has been derived. The order of the eight benchmark suites in terms of the pipeline stall degree matches well with the order of their energy level curves.

%\section{Discussion}
\section{Conclusions}
\label{sec:Conclusions}

As computers are widely applied in increasingly more areas, fixed computer architecture contradicts the increasing diversity of applications, resulting in the utilization wall and the memory wall. In this study, we proposed a Recursive Reduce methodology to obtain the Gene-Patterns of diverse applications. This methodology can be conducted in commercial processors based on performance counters, thus it is much faster than the methods based on simulators. Our study showed that there exist significant differences among the applications in different domains, but there also exist many similarities from a pattern perspective, which makes the amount of base patterns very limited. We have identified a set of six base patterns. We found that inefficiency results from the mismatch between some of the base patterns and the micro-architecture of modern processors. We build a periodic table, where the indifference curves are analogous to the energy levels in physics. The location in the table quantifies the matching degree between a pattern and an architecture, and memory performance optimization is essentially an energy level transition.

As we have already found the ``genes'' of applications (when thinking of applications as ``lives''), we could take these genes as building blocks for the design of both architecture and software. It is not necessary to develop customized architecture for each application. The findings in our study will facilitate accelerator design, and hybrid or heterogeneous system design. These Gene-patterns will straddle the divide between general-purpose and special purpose, providing a methodology for the matching between application diversity and architecture fixity. 

\bibliographystyle{unsrt}
\bibliography{references}

\end{document}